\documentstyle[twoside]{article}

\catcode`\@=11
\long\def\@makefntext#1{
\protect\noindent \hbox to 3.2pt {\hskip-.9pt  
$^{{\eightrm\@thefnmark}}$\hfil}#1\hfill}		%CAN BE USED 

\def\@makefnmark{\hbox to 0pt{$^{\@thefnmark}$\hss}}	%ORIGINAL 
	
\def\ps@myheadings{\let\@mkboth\@gobbletwo
\def\@oddhead{\hbox{}
\rightmark\hfil\eightrm\thepage}   
\def\@oddfoot{}\def\@evenhead{\eightrm\thepage\hfil
\leftmark\hbox{}}\def\@evenfoot{}
\def\sectionmark##1{}\def\subsectionmark##1{}}

\oddsidemargin=\evensidemargin
\newcounter{sectionc}\newcounter{subsectionc}\newcounter{subsubsectionc}
\renewcommand{\section}[1] {\vspace{12pt}\addtocounter{sectionc}{1} 
\setcounter{subsectionc}{0}\setcounter{subsubsectionc}{0}\noindent 
	{\tenbf\thesectionc. #1}\par\vspace{5pt}}
\renewcommand{\subsection}[1] {\vspace{12pt}\addtocounter{subsectionc}{1} 
	\setcounter{subsubsectionc}{0}\noindent 
	{\bf\thesectionc.\thesubsectionc. {\kern1pt \bfit #1}}\par\vspace{5pt}}
\renewcommand{\subsubsection}[1] {\vspace{12pt}\addtocounter{subsubsectionc}{1}
	\noindent{\tenrm\thesectionc.\thesubsectionc.\thesubsubsectionc.
	{\kern1pt \tenit #1}}\par\vspace{5pt}}
\newcommand{\nonumsection}[1] {\vspace{12pt}\noindent{\tenbf #1}
	\par\vspace{5pt}}

\newcommand{\textlineskip}{\baselineskip=13pt}
\newcommand{\smalllineskip}{\baselineskip=10pt}

\def\eightcirc{
\begin{picture}(0,0)
\put(4.4,1.8){\circle{6.5}}
\end{picture}}
\def\eightcopyright{\eightcirc\kern2.7pt\hbox{\eightrm c}} 

\newcommand{\copyrightheading}[1]
{\vspace*{-2.5cm}\smalllineskip{\flushleft
     {\footnotesize {\bf Quantum Aspects of Beam Physics 3}}\\ 
      {\footnotesize Higashi-Hiroshima, Japan,
      7-11 January 2003}\\
      {\footnotesize Ed. Pisin Chen, pp. --- #1}\\
      {\footnotesize $\eightcopyright$\ 2003 by World Scientific Publishing
        Co. Pte. Ltd.}\\
	 }}

\def\abstracts#1#2#3{{
	\centering{\begin{minipage}{4.5in}\baselineskip=10pt\footnotesize
	\parindent=0pt #1\par 
	\parindent=15pt #2\par
	\parindent=15pt #3
	\end{minipage}}\par}} 

\renewenvironment{thebibliography}[1]
	{\frenchspacing
	 \ninerm\baselineskip=11pt
	 \begin{list}{\arabic{enumi}.}
        {\usecounter{enumi}\setlength{\parsep}{0pt}     
	 \setlength{\leftmargin 12.7pt}{\rightmargin 0pt} %FOR 1--9 ITEMS
         \setlength{\itemsep}{0pt} \settowidth
	{\labelwidth}{#1.}\sloppy}}{\end{list}}

\newcounter{itemlistc}
\newcounter{romanlistc}
\newcounter{alphlistc}
\newcounter{arabiclistc}

\def\@citex[#1]#2{\if@filesw\immediate\write\@auxout
	{\string\citation{#2}}\fi
\def\@citea{}\@cite{\@for\@citeb:=#2\do
	{\@citea\def\@citea{,}\@ifundefined
	{b@\@citeb}{{\bf ?}\@warning
	{Citation `\@citeb' on page \thepage \space undefined}}
	{\csname b@\@citeb\endcsname}}}{#1}}

\newif\if@cghi
\def\cite{\@cghitrue\@ifnextchar [{\@tempswatrue
	\@citex}{\@tempswafalse\@citex[]}}
\def\citelow{\@cghifalse\@ifnextchar [{\@tempswatrue
	\@citex}{\@tempswafalse\@citex[]}}
\def\@cite#1#2{{$\null^{#1}$\if@tempswa\typeout
	{IJCGA warning: optional citation argument 
	ignored: `#2'} \fi}}

\def\@refcitex[#1]#2{\if@filesw\immediate\write\@auxout
	{\string\citation{#2}}\fi
\def\@citea{}\@refcite{\@for\@citeb:=#2\do
	{\@citea\def\@citea{, }\@ifundefined
	{b@\@citeb}{{\bf ?}\@warning
	{Citation `\@citeb' on page \thepage \space undefined}}
	\hbox{\csname b@\@citeb\endcsname}}}{#1}}

\def\@refcite#1#2{{#1\if@tempswa\typeout
        {IJCGA warning: optional citation argument
	ignored: `#2'} \fi}}

\def\refcite{\@ifnextchar[{\@tempswatrue
	\@refcitex}{\@tempswafalse\@refcitex[]}}

\def\pmb#1{\setbox0=\hbox{#1}
	\kern-.025em\copy0\kern-\wd0
	\kern.05em\copy0\kern-\wd0
	\kern-.025em\raise.0433em\box0}

\def\fnt#1#2{\footnotetext{\kern-.3em
	{$^{\mbox{\scriptsize #1}}$}{#2}}}

\font\tenrm=cmr10
\font\tenit=cmti10 
\font\tenbf=cmbx10
\font\bfit=cmbxti10 at 10pt
\font\ninerm=cmr9

\font\eightrm=cmr8

\def\qed{\hbox{${\vcenter{\vbox{			%HOLLOW SQUARE
   \hrule height 0.4pt\hbox{\vrule width 0.4pt height 6pt
   \kern5pt\vrule width 0.4pt}\hrule height 0.4pt}}}$}}

\begin{document}

%\runninghead{Instructions for Typesetting Camera-Ready
%Manuscripts $\ldots$} {Instructions for Typesetting Camera-Ready
%Manuscripts $\ldots$}

%Comment (H. Rosu): produce fraza de mai sus la inceputul fiecarei pagini

\normalsize\textlineskip
\setcounter{page}{1}

\copyrightheading{}			%{Vol. 0, No.0 (1992) 000--000}

\vspace*{0.88truein}

%\fpage{1} %%%%%%%%%%%%%%%%%%%%%%%%%%%%%%%%%%%%%%%%%%%%%%%%%%%%%%%%%%%
\centerline{\bf {UNRUH EFFECT AS PARTICULAR FRENET SERRET VACUUM RADIATION}} 
\medskip
\centerline{\bf {AND DETECTION PROPOSALS}}
\vspace*{0.035truein}
%\centerline{\bf MANUSCRIPTS USING COMPUTER SOFTWARE\footnote{For
%the title, try not to use more than 3 lines. Typeset the title
%in 10 pt Times Roman, uppercase and boldface.}}
\vspace*{0.37truein}
\centerline{\footnotesize HARET C. ROSU} 
\centerline{\footnotesize hcr@ipicyt.edu.mx}
%\footnote{Typeset names in
%10 pt Times Roman, uppercase. Use the footnote to indicate the
%present or permanent address of the author.}}
\vspace*{0.015truein}
\centerline{\footnotesize\it Applied Mathematics and Computational Systems}
\centerline{\footnotesize \it Potosinian Institute of Scientific and Technological Research}
\centerline{\footnotesize \it Apdo Postal 3-74 Tangamanga, San Luis Potos\'{\i}, SLP, Mexico}
\baselineskip=10pt
%\centerline{\footnotesize\it City, State ZIP/Zone,
%Country\footnote{State completely without abbreviations, the
%affiliation and mailing address, including country. Typeset in 8
%pt Times Italic.}}
\vspace*{10pt}
%\centerline{\footnotesize SECOND AUTHOR}
%\vspace*{0.015truein}
%\centerline{\footnotesize\it Group, Laboratory, Address}
%\baselineskip=10pt
%\centerline{\footnotesize\it City, State ZIP/Zone, Country}
\vspace*{0.225truein}
%\publisher{(May 27, 1997)}{(December 29, 1997)}

\vspace*{0.21truein}
\abstracts{The paradigmatic Unruh radiation is an ideal and simple case of stationary scalar vacuum radiation patterns related to worldlines defined
as Frenet-Serret curves. We briefly review the corresponding body of theoretical literature as well as the proposals that have been suggested
to detect these types of quantum field radiation patterns. %Finally, we comment on a few other topics related to the Unruh effect.
}{}{}

%\vspace*{10pt}
%\keywords{The contents of the keywords}

\textlineskip                  %) USE THIS MEASUREMENT WHEN THERE IS
\vspace*{9pt}                 %) NO SECTION HEADING

\vspace*{1pt}\textlineskip	%) USE THIS MEASUREMENT WHEN THERE IS
%\section{General Appearance}    %) A SECTION HEADING
\vspace*{-0.5pt}
\noindent

%%%%%%%%%%%%%%%%%%%%%%%%%%%%%%%%%%%%%%%%%%%%%%
%PACS number(s):  98.80.Hw, 11.30.Pb
%\vskip 2cm

\noindent
%%%%%%%%%%%%%%%%%%%%%%%%%%%%%%%%%%%%%%%%%%%%%%%%%%%%%%%%%%%%%%%%%%%%%

%\newpage

%\pagebreak

%\textheight=7.8truein
%\setcounter{footnote}{0}
%\renewcommand{\thefootnote}{\alph{footnote}}

%\section{The Main Text}
\noindent

% 1.  %%%%%%%%%%%%%%%%%%%%%%%%%%%%%%%%%%%%%%%%%%%%%%%%%%%%%%%%%%%%%%%%%%%%
\section{Frenet-Serret Worldlines and Vacuum Radiation Patterns}
%%%%%%%%%%%%%%%%%%%%%%%%%%%%%%%%%%%%%%%%%%%%%%%%%%%%%%%%%%%%%%%%%%%%%%%%%
%\begin{itemize}
%\item 1.
%\end{itemize}

\noindent
A thermal radiation effect due to vacuum oscillations in quantum field theory has been discussed by Unruh in 1976,\cite{u76} using the so-called {\em detector method}.
This was based on the first order perturbation calculation of the excitation rate of a quantum particle considered as a two-level field detector around its classical trajectory.
Slightly earlier, Davies obtained a similar result using a 
{\em mirror model},\cite{dav75} that implies the calculation of the Bogoliubov coefficient $\beta$, like in particle production in astrophysics and cosmology. A `thermodynamic' temperature, $T_V=\frac{\hbar}{2\pi ck_B}\cdot a$, directly proportional to the proper linear acceleration $a$ is the main feature of this vacuum radiation pointing to a new universal quantum field thermal effect. Moreover, a direct link to the Hawking radiation in black hole physics could be thought of through the equivalence principle. 
On the other hand, in a little noticed paper of 1981, Letaw \cite{l81} studied by means of Frenet-Serret tetrads and the same detector method the  
stationary world lines on which relativistic quantum particles with a linear coupling to the scalar vacuum have time-independent excitation
spectra. These worldlines are characterized by the requirement that the geodetic interval between two points depends only
on the proper time interval. Letaw employed a generalization of the Frenet-Serret equations to the four-dimensional Minkowski space in which the worldlines are 
characterized by the curvature $\kappa$ and two torsions $\tau _1$ and $\tau _2$ instead of a single one as in the common three-dimensional space. Mathematically, this means a change of dimension of the antisymmetric matrix of curvature invariants
%%%%%%%%%%%%%%%
\begin{equation}
\left(\begin{array}{ccc}
 0 & \kappa & 0 \\
-\kappa & 0 & \tau\\
0 & -\tau & 0\end{array} \right )\Longrightarrow
\left(\begin{array}{cccc}
 0 & -\kappa (s)& 0 & 0\\
\kappa (s)& 0 & -\tau _1 (s)& 0\\
0 & \tau _1(s) & 0 & -\tau _2(s)\\
0 & 0 & \tau _ 2(s) & 0 \end{array} \right )~,
\end{equation}
where $s$ is the proper time parameter along the classical Frenet-Serret trajectory.
Not surprisingly, the curvature invariants are the proper acceleration and angular velocity of the world line. Solving the generalized Frenet-Serret equations for the simple case of
constant invariants leads to  six classes of stationary world lines. He also demonstrated the
equivalence of the timelike Killing vector field orbits and the stationary world lines. 
%The classification scheme therefore extends to Killing orbits and stationary coordinate systems in flat spacetime. 
Last but not least, Letaw did some calculations of the vacuum excitation spectra of detectors on the sample of six families of stationary world lines, i.e., of the following 
cosine Fourier transform
%%%%%%%%%%%%%%%
\begin{equation} \label{l81-04}
S(E, \tau)=2\pi \rho (E) \int _{-\infty}^{0}ds\langle 0|\phi(x(\tau))\phi(x(\tau +s))|0\rangle\cos(Es)~,
\end{equation}
%%%%%%%%%%%%%%%
$E$ is the energy difference between the two levels of the particle considered as detector
of the vacuum spectra, $\rho(E)$ is the density of states of the detected vacuum `quasiparticles', and $\langle 0|\phi(x(\tau))\phi(x(\tau +s))|0\rangle$ is the expectation value of the Wightman autocorrelation function in the ground state of the 
particle.
 %is time independent the detected spectra are stationary.
Letaw's work is a generalization of Unruh's result concerning the excitation of a scalar particle detector moving with constant linear acceleration in the vacuum of flat spacetime. Unruh's result became famous because of
Unruh's interpretation that the detector behaves as if in contact with a bath of scalar `particles' with energies in a Planck spectrum of temperature proportional to $a/2\pi$
($\hbar\,,c\,,k_B=1$). 
The connection with the Hawking radiation and its paradigmatic nature  led many theoretical physicists to focus on Unruh's effect and there is a strong need for an
experimental confirmation of the effect as a consequence of long debate.\cite{bd} It is the main goal of this short survey to present the ideas that have been generated over
the years in this respect.

\medskip

\subsection{The Six Stationary Scalar Frenet-Serret Radiation Spectra}

\noindent
We quote here those vacuum excitation spectra $S(E,\tau)$ that are independent of proper time $\tau$, i.e., stationary.

\medskip
\noindent
\underline{{\bf 1}. Inertial (uncurved) worldlines} $\quad$ $\kappa =\tau _1=\tau _2=0$ $\quad$ %$\rightarrow$$\quad$
%%%%%%%%%%%%%%%%%%%%%%%%%%%%%%%%%%%%%%%%%%%%%%%%%%%%%%%%%%%%%%%%%%%%%%%%%%%%%%%
%\bigskip
%\noindent
%The excitation spectrum is a trivial cubic spectrum 
%\bigskip
%%%%%% 
%\begin{equation} 

\medskip

$S_{0}(E)=\frac{E^3}{4\pi ^2}$~.
%\end{equation} 
%%%%%%%%% 

\medskip

\noindent
The interpretation is a normal vacuum spectrum, i.e., as given by a vacuum of zero point energy per mode $E/2$ 
and density of states $E^2/2\pi ^2$.

\bigskip 
\noindent 
\underline{{\bf 2}. Hyperbolic worldlines} $\quad$ $\kappa \neq 0$, $\tau _1=\tau _2=0$ %$\quad$ $\rightarrow$
%%%%%%%%%%%%%%%%%%%%%%%%%%%%%%%%%%%%%%%%%%%%%%%%%%%%%%%%%%%%%%%%%%%%%%%%%%
$\quad$ 

\medskip

$S_{\kappa}(\epsilon _{\kappa}) 
=\frac{\epsilon _{\kappa}^{3}}{2\pi ^2(e^{2\pi\epsilon _{\kappa}}-1)}$~.

\bigskip
\noindent
This is the unique noninertial case that is torsionless. The employed
variable is $\epsilon _{\kappa}=E/\kappa$.The excitation spectrum is Planckian allowing the 
interpretation of $\kappa/2\pi$ as `thermodynamic' temperature. 

\bigskip 
\noindent 
\underline{{\bf 3}. Ultratorsional (helical) worldlines} $\quad$$|\kappa|<|\tau _1|\neq 0$, $\tau _2=0$, $\rho ^2=\tau _1 ^2-\kappa ^2$ %$\rightarrow$
%%%%%%%%%%%%%%%%%%%%%%%%%%%%%%%%%%%%%%%%%%%%%%%%%%%%%%%%%%%%%%%%%%%%%%%%%%%%%%%%%%%%%%%%%%%%

\medskip

$S^{{\rm u}}_{\tau _1}(\epsilon _{\rho})\stackrel{\kappa/\rho\rightarrow \infty} 
{\longrightarrow} S^{{\rm p}}_{ \tau _1}(\epsilon _{\kappa})$~.

\bigskip
\noindent
The excitation spectrum is an analytic function corresponding to the case 4 below only in the limit $\kappa\gg \rho$. 
%%%%%%%%% 
Letaw plotted the numerical integral for $S_{\tau _1}^{\rm u}(\epsilon _{\rho})$, 
where $\epsilon _{\rho}=E/\rho$ for various values of $\kappa/\rho$.

\bigskip 
\noindent 
\underline{{\bf 4}. Paratorsional (semicubical parabolic) worldlines}$\quad$ $\kappa=\tau _1\neq 0$, $\tau _2=0$ %$\quad$ $\rightarrow$
%%%%%%%%%%%%%%%%%%%%%%%%%%%%%%%%%%%%%%%%%%%%%%%%%%%%%%%%%%%%%%%%%%%%%%%%%%%%%%%%%%%%% 

\medskip

$S^{{\rm p}}_{\tau _1}(\epsilon _{\kappa})= \frac{\epsilon _{\kappa}^{2}}{8\pi ^2 \sqrt{3}} 
e^{-2\sqrt{3}\epsilon _{\kappa}}~.$%(if spatially projected: semicubical parabolas $y=\frac{\sqrt{2}}{3}\kappa x^{3/2}$) 

\medskip
%(the spatially projected worldlines are the semicubical parabolas 
%$y=\frac{\sqrt{2}}{3}\kappa x^{3/2}$ 
%This parabolas contain a cusp where the direction of motion is reversed. 

\noindent
The excitation spectrum is analytic, and 
since there are two equal curvature invariants one can use the 
dimensionless energy variable $\epsilon _{\kappa}$. 
%(for mathematical details see the Appendix) 
%but is not Planckian)
It is worth noting that $S^{{\rm p}}_{\tau _1}$, being a monomial times an exponential, 
is quite close to the Wien-type spectrum 
$S_{W}\propto\epsilon ^3e^{- {\rm const.}\epsilon}$.

\bigskip 
\noindent 
\underline{{\bf 5}.  Infratorsional (catenary) worldlines} $\quad$ $|\kappa|>|\tau _1|\neq 0$, $\tau _2=0$, $\sigma ^2=\kappa ^2-\tau _1 ^2$ 
%%%%%%%%%%%%%%%%%%%%%%%%%%%%%%%%%%%%%%%%%%%%%%%%%%%%%%%%%%%%%%%%%%%%%%%%%%%%%%%%%%%%%%%

\medskip

$S_{\kappa}(\epsilon _{\kappa}) 
\stackrel{0\leftarrow \tau/\sigma}{\longleftarrow} 
S^{{\rm i}}_{\tau _1}(\epsilon _{\sigma})\stackrel{\tau/\sigma\rightarrow \infty} 
{\longrightarrow}S^{{\rm p}}_{\tau _1}(\epsilon _{\kappa})$.
%$x=\kappa \cosh (y/\tau)$) 

\medskip
%(the spatially projected worldlines are catenaries, i.e., curves of the type 
%$x=\kappa \cosh (y/\tau)$). 

\noindent
In general, the catenary spectrum cannot be found 
analytically. It is an intermediate case, which 
for $\tau/\sigma\rightarrow 0$ tends to $S_{\kappa}$, 
whereas for $\tau/\sigma\rightarrow\infty$ tends toward $S^{{\rm p}}_{\tau _1}$. 

%\bigskip
%%%%%%%%%%%%%%%%%%%%%%%%%%%%% 
%\begin{equation} 
%S_{\rm hyp}(\epsilon _{\kappa}) 
%\stackrel{0\leftarrow \tau/\sigma}{\longleftarrow} 
%S_{\rm catenary}(\epsilon _{\sigma})\stackrel{\tau/\sigma\rightarrow \infty} 
%{\longrightarrow}S_{\rm 3/2-parab}(\epsilon _{\kappa})~. 
%\end{equation} 
%%%%%%%%%%%%%%%%%%%%%%%%%%%

\medskip 
\noindent 
\underline{{\bf 6}. Hypertorsional (variable pitch helicoid) worldlines}$\quad$ $\tau _2 \neq 0$ 

\medskip 
$S_{\tau _2}$ is not analytic.

\medskip

\noindent
The hypertorsional worldlines are rotating with constant $a_{\perp}$ to the rotation plane. 
The excitation spectrum is given in this case by a two-parameter set of curves. These trajectories are 
a superposition of the constant linearly accelerated motion and uniform 
circular motion. According to Letaw, the spatial path of a two-level detector on this world line is helicoid of variable pitch that decreases to zero at 
proper time interval $\tau =0$ and increases thereafter. The corresponding vacuum spectra have not been calculated 
by Letaw, not even numerically.

\medskip

\subsection{Conclusions from the stationary scalar cases}

\noindent
Examining the six scalar stationary cases we see that only the hyperbolic worldlines, having just one nonzero curvature 
invariant, allow for a Planckian excitation spectrum and lead to a strictly one-to-one 
mapping between the curvature invariant $\kappa$ and the `thermodynamic' 
temperature ($T_{\kappa}=T_{V}=\kappa /2\pi$). The excitation spectrum due to  semicubical 
parabolas can be fitted by Wien type spectra, the radiometric parameter 
corresponding to both curvature and torsion. 
The other stationary cases, being nonanalytical, lead to 
approximate determination of the curvature invariants, defining locally the 
classical worldline on which a relativistic quantum particle moves. This explains 
why the Unruh effect became so prominent with regard to the other five types of stationary Frenet-Serret scalar spectra.

\medskip

\noindent
For the important case of electromagnetic vacuum fluctuations the FS formalism has not been used in a direct way. However, Hacyan and Sarmiento \cite{hs}
developed a procedure by which they provided nonanalytic formulas (cosine Fourier transform integrals) 
for the spectral energy density, flux density, and stress density of the vacuum radiation
in terms of the electromagnetic Wightman functions calculated by means of the two Killing vectors associated to circular trajectories. 
%Although the results in terms of the Killing vectors should be equivalent to the FS calculation the comparison is not a direct one.

\section{\bf  Detection Proposals}

\medskip
\noindent
%A number of model experiments to detect `Unruh radiation' have been suggested in the last twenty years. The author has undertaken 
 %the task of reviewing most of them \cite{rev1,rev2}.
Because the curvature thermodynamic temperature is given by $T_{\kappa}=\frac{\hbar}{2\pi c k}\,a$ this leads to $T_{\kappa}=4\cdot 10^{-23}\, a$
and one needs accelerations greater than $10^{20}g_{\oplus}$ to have `thermal' effects of only a few Kelvin degrees. On the other hand, one should focus
%There are indeed several physical settings in which accelerations only a few orders 
below the Schwinger acceleration for copious spontaneous 
pair creation out of QED vacuum, ${a}_{\rm Schw} \approx m_e c^3/\hbar \approx 10^{29} \; m/s^2 \approx 10^{28} \; g_{\oplus}$.
Thus the optimal range for detecting a possible Unruh effect entails eight orders of magnitude in proper acceleration
%%%%%%%%%%%%%%%%%%%%%
\begin{equation} \label{accel}
10^{20}g_{\oplus}\leq a \leq 10^{28} \; g_{\oplus}~,
\end{equation}
%%%%%%%%%%%%%%%%%%%%
 There are indeed several physical settings (for reviews, see \cite{reviews}) in which accelerations can be achieved only a few orders 
below the Schwinger acceleration and forthcoming technological advances could test routinely those acceleration scales. The Unruh effect, if it exists,
can be revealed as a tiny thermal-like signal in the background of by far more powerful effects.

The following is the list of proposals.
\medskip

%\begin{enumerate}

%\item 

\subsection{Unruh Effect in Storage Rings ($a\sim 10^{22}g_{\oplus}, \, T_{\kappa}=1200\, K$)}
%%%%%%%%%%%%%%%%%%%%%%%%%%%%%%%%%%%%%%%%%%%%%%%%%%%%%%%%%%%%

\medskip

\noindent
J.S. Bell and J.M. Leinaas imagined the first laboratory phenomenon connected to the
Unruh effect. During 1983-1987 they published a number of papers on the idea that the depolarising effects in
electron storage rings could be interpreted in terms of Unruh
effect.\cite{bellLei}%,mcd1} % see also \cite{mcd1}).
However, the incomplete radiative polarization of the electrons in storage rings has been first predicted in early sixties by Sokolov and Ternov,\cite{sokte}
as an effect due to the spin-flip synchrotron radiation in the framework of QED. Their approach successfully provides the observed maximum polarization of 
electrons at storage rings, $P_{max}=\frac{8\sqrt{3}}{15}=0.924$.\cite{spear} Besides, the circular vacuum noise is not sufficiently ``universal" since it
always depends on both acceleration and velocity.
%does not have the same universal thermal character as the linear Unruh noise.\cite{tak} 
This appears as a `drawback' of the {\em storage ring electron radiometry},\cite{ku} not to
mention the very intricate spin physics.

The polarization calculated by Bell and Leinaas is very similar in shape to a formula for the polarization as a function of the electron gyromagnetic factor $g$
obtained by Derbenev and Kondratenko,\cite{der} in 1973 that is considered the standard QED accelerator result for the polarization of beams. Their 
function $ P_{DK}(g)$ is a combination of exponential and polynomial terms in the anomalous part of the gyromagnetic factor of the electron.
Barber and Mane \cite{bar1} have shown that the DK and BL formalisms for the equilibrium degree of radiative electron
polarization are not so different as they might look. They also obtained an even more general formula for the equilibrium polarization
than the DK and BL ones and from their formula they estimated as negligible the differences between them. %BL increase near the resonance.

Recently, the spin-flip synchrotron radiation has been experimentally shown to be important
in the hard part of the spectrum in the axial channeling of electrons in the energy range
35-243 GeV incident on a W single crystal.\cite{kirsebom} This may revive the interest
in the BL interpretation, especially in the cleaner planar channeling case.\cite{infou} 

One can also recall that K.T. McDonald applied the Unruh temperature formula for a rapid calculation of the damping in a linear focusing channel.\cite{kirk} This is a transport 
system at accelerators that confines the motion of charged particles along straight central rays by means of a potential quadratic in the transverse spatial coordinates.
He used the same idea about two decades ago to reproduce Sands' results on the limits of damping of the phase volume of beams in electron storage rings.

\medskip

%\begin{enumerate}

\subsection{Unruh Effect and the Physics of Traps  ($a\sim 10^{21}g_{\oplus}, \, T_{\kappa}=2.4\,K$)}
%%%%%%%%%%%%%%%%%%%%%%%%%%%%%%%%%%%%%%%%%%%%%%%%%%%%%%%%%%%%

\medskip

\noindent
The very successful and precise physics of traps could help detecting the circular
thermal-like vacuum noise. The proposal belongs to J. Rogers \cite{rog}
being one of the most attractive.
The idea of Rogers is to place a small superconducting Penning trap in a microwave cavity. A single
electron is constrained to move in a cyclotron orbit around the trap
axis by a uniform magnetic field (Rogers' figure is B = 150 kGs).
The circular proper acceleration is $a= 6\times 10^{21}g_{\oplus}$
corresponding to T = 2.4 K. The velocity of the electron is maintained fixed
($\beta= 0.6$) by means of a circularly polarized wave at the electron
cyclotron frequency, compensating also for the irradiated power.
The static quadrupole electric field of the trap creates a quadratic
potential well along the trap axis in which the electron oscillates.
The axial frequency is 10.5 GHz (more than 150 times the typical
experimental situation \cite{bg}) for the device scale chosen by
Rogers. This is the measured frequency since it is known that the best way
of observing the electron motion from the outside world is through the measurement of 
the current due to the induced charge on the cap electrodes of the trap, as a consequence of the axial
motion of the electron along the symmetry axis.\cite{bg} At 10.5 GHz the difference in energy densities between the circular
electromagnetic vacuum noise and the universal linear scalar noise are negligible (see Fig. 2 in Rogers' work).
Even better experimental setups in this context could be electrons in cylindrical
Penning traps with the trap itself representing the microwave cavity.\cite{cylP}

\medskip

\subsection{Unruh Effect and Nonadiabatic Casimir Effect  ($a\sim 10^{20}g_{\oplus},\, T\sim 1\, K$)}
%%%%%%%%%%%%%%%%%%%%%%%%%%%%%%%%%%%%%%%%%%%%%%%%%%%%%%%%%%%%%%%

\medskip\noindent
Yablonovitch,\cite{yab} proposed a {\em plasma front} as 
an experimental equivalent of a {\em fast moving mirror}.
Plasma fronts can be created when a gas is suddenly photoionized. 
The argument is that the phase shift of the zero-point electromagnetic field
transmitted through a plasma window whose index of refraction
is falling with time (from 1 to 0) is the same as when reflected
from an accelerating mirror. Consider the case of hyperbolic motion.
Since the velocity is
%%%%%%%%%%%%%%%%%%%%%%%
\begin{equation}\label{yab1}
v= c\tanh(a\tau/c)
\end{equation}
%%%%%%%%%%%%%%%%%%%%%
where $\tau$ is the observer's proper time, the Doppler shift frequency
will be
%%%%%%%%%%%%%%%%
\begin{equation}\label{yab2}
\omega_{D} = \omega_{0}\sqrt{\frac{1 - v/c}{1 +v/c}} =
\omega_{0}\exp({-a\tau/c})
\end{equation}
%%%%%%%%%%%%%%%%
and consequently a plane wave of frequency $\omega_{0}$ turns into
a wave with a time-dependent frequency. Such waves are called chirped
waves in nonlinear optics and acoustics. Eq. (\ref{yab2}) represents an
{\em exponential chirping} valid also for Schwartzschild black holes with 
the substitution $a=c^4/4GM$ ($G$ is Newton's constant and $M$ is the mass 
parameter of the Schwarzschild black hole). 

The technique of producing plasma fronts/windows in a gas by laser breakdown,
and the associated frequency upshifting phenomena (there are also downshifts)
of the electromagnetic waves interacting with such windows,
are well settled since about twenty years.
Blue shifts of about $10\%$ have been usually observed
in the transmitted laser photon energy.

In his paper, Yablonovitch works out a very simple model of a {\em linear}
chirping due to a refractive index linearly decreasing with time,
$n(t)=n_{0}-\dot n t$, implying a Doppler shift of the form
$\omega \rightarrow \omega[1+\frac{\dot n}{n} t]\sim \omega[1+\frac{a}{c} t]$.
To have accelerations $a =10^{20}g_{\oplus}$ the laser pulses should be
less than 1 picosecond.
Even more promising may be the nonadiabatic photoionization
of a semiconductor crystal in which case the refractive index
can be reduced from 3.5 to 0 on the timescale of the optical
pulse. As discussed by Yablonovitch, the pump laser has to be tuned
just below the Urbach tail of a direct-gap semiconductor in order
to create weakly bound virtual electron-hole pairs. These pairs contribute a large
reactive component to the photocurrent since they are readily polarized. The
background is due to the bremsstrahlung emission produced by real electron-hole pairs, and to 
diminish it one needs a crystal with a big Urbach slope (the Urbach tail is an
exponential behavior of the absorption coefficient).

\medskip
\noindent
In addition, Eberlein,\cite{eb1} elaborated on Schwinger's interpretation of sonoluminescence in terms of zero point fluctuations and asserted 
that whenever {\em an interface between two dielectrics or a dielectric and the vacuum moves noninertially photons are created}, i.e., the 
Unruh effect occurs. An interesting discussion in favor of ``dielectric windows" rather than the ``plasma window" is provided by Dodonov {\em et al}.\cite{dwin}
Moreover, Grishchuk, Haus, and Bergman,\cite{ghb92} discussed a nonlinear Mach-Zhender configuration to generate radiation 
through the optical squeezing of zero-point fluctuations interacting with a {\em moving index grating} that is also reminiscent of Unruh effect. 

\medskip

\subsection{Unruh Effect and Channeling ($a\sim 10^{30}g_{\oplus},\, T_{\kappa} \sim 10^{11}\, K\, ?$)}
%%%%%%%%%%%%%%%%%%%%%%%%%%%%%%%%%%%%%%%%%%%%%%%%%%%%%%%%%%%%%%%%%

\medskip
\noindent
Relativistic particles can acquire extremely high transverse accelerations when they are channeled through crystals. 
Darbinian and collaborators \cite{dar} related this physical setting to Unruh radiation.

The idea is to measure the {\em Unruh radiation emitted in the Compton scattering 
of the channeled particles with the Planck spectrum of the inertial crystal vacuum}. The main argument
is that the crystallographic fields act with large transverse accelerations on the channeled particles.
The estimated transverse proper acceleration for positrons
channeled in the (110) plane of a diamond crystal is $a = 10^{25}\gamma$ cm/s$^{2}$, and
at a $\gamma = 10^{8}$ one could reach $10^{33}$ cm/s$^{2}$ = $ 10^{30}g_{\oplus}$.
Working first in the particle instantaneous rest frame, Darbinian {\em et al} derived the spectral angular distribution of the Unruh photons
in that frame. By Lorentz transformation to the lab system they got the number of Unruh photons per unit length of crystal and averaged over
the channeling diameter. At about $\gamma = 10^{8}$ the Unruh intensity, i.e., the intensity per unit pathlength of the Compton scattering on the
Planck vacuum spectrum becomes comparable with the Bethe-Heitler bremsstrahlung ($dN_{\gamma}/dE\propto 1/E$, and mean polar emission 
angle $\theta =1/\gamma$).

%\medskip
Similar calculations have been applied by the same group,\cite{dar1} to get an estimate of the Unruh radiation generated by TeV electrons in a 
uniform magnetic field as well as in a circularly polarized laser field but the conclusions are not optimistic because of the huge synchrotron background. 

%K.T. McDonald applied the Unruh temperature formula for a rapid calculation of the damping in a linear focusing channel. This is a transport 
%system that confines the motion of charged particles along straight central rays by means of a potential quadratic in the transverse spatial coordinates.
%He used the same idea about two decades ago to reproduce Sands' results on the limits of damping of the phase volume of beams in electron storage rings. 

\subsection{Unruh Radiation and Ultraintense Lasers ($a\sim 10^{25}g_{\oplus},\,T_{\kappa}= 1.2\, 10^{6}\,K$)}
%%%%%%%%%%%%%%%%%%%%%%%%%%%%%%%%%%%%%%%%%%%%%%%%%%%%%%%%%%%%%%%%%%%%%%

\medskip

\noindent
A Unruh signal could be obtained in electron Petawatt-class laser interaction according to a proposal put forth by Chen and Tajima in 1999.\cite{ct99}
Uniform acceleration through the usual quantum vacuum 
(Minkowski vacuum) of the electromagnetic field distorts the 
two-point function of the zero-point fluctuations in such a way that 
\begin{equation}\label{viss1}
\langle E_{i}(-\tau/2) E_{j}(+\tau/2) \rangle = 
{4\hbar\over\pi c^3} \; \delta_{ij} \; 
{(a/c)^4 \over \sinh^4(a \tau/2c)}~.
\hfill\qquad\qquad  
\end{equation}

The main point of Tajima and Chen is to introduce the so-called laser strength (ponderomotive) parameter $a_0=\frac{eE_0}{mc\omega _0}$ in this formula and 
in all their estimations. They calculate the Unruh radiation based on the autocorrelation function in Eq.~(\ref{viss1}).
The accelerated electron is assumed ``classical", i.e., with well-defined acceleration, velocity, and position.
This allows to introduce a Lorentz transformation so that the electron is described in its instantaneous proper frame. In the words of Chen and 
Tajima ``the electron reacts to the vacuum fluctuations with 
a nonrelativistic quivering motion in its proper frame" that triggers additional (Unruh) radiation besides the classical Larmor radiation.

The important claim of Chen and Tajima is that there is a blind spot in the Larmor angular distribution for azimuthal angle $\Delta \phi =10^{-3}$ and polar angle $\Delta \theta \ll 1/a_0$ where the Unruh thermal-like signal could be revealed.  
Since at each half cycle the electron almost suddenly becomes relativistic, with constant $\gamma \sim a_0$, the Unruh radiation is boosted along the direction of 
polarization in the lab frame. Moreover, they showed that the autocorrelation function, and therefore the Unruh signal, tend to 
diminish more rapidly than that from Larmor within the laser half cycle. This should induce a sharper time structure for the former that could help its detection.

%\medskip

\subsection{Unruh Radiation in Quantum Optics {\it (moderate $a$ could work)}}

\noindent
This is a very recent proposal in several versions due to a Scully collaboration.\cite{s03}
The idea is to enhance the thermal Unruh radiation signal from an accelerating  He$^+$ ion used as a two-level type detector of transition frequency $\omega$
passing through a high $Q$ 
``single mode" cavity of frequency $\nu$ in the vicinity of the atomic frequency $\omega$. The enhancement is very significant, in the sense that for reasonable values of the parameters, the effective Boltzmann factor turns
from the usual exponential behaviour to a linear dependence in $\alpha/2\pi \omega$, where $\alpha = a/c$.
Employing quantum optics calculations, they showed that this type of Unruh effect is due to nonadiabatic transitions stemming from the counter-rotating 
term $\hat{a}_{k}^{+}\hat{\sigma}^{+}$ in the time-dependent atom-field interaction Hamiltonian. The Larmor radiation lobes ($\sim \sin ^2 \theta$) 
will certainly be present but the blind spot in the forward direction of motion could be hopefully used for the detection of this nonadiabatic thermal effect. 

%\end{enumerate}

\medskip

\section{Conclusion}

\medskip
\noindent
Although the Unruh radiative effect is interpreted as a thermal effect of the noninertially and nonadiabatically-produced vacuum state, its thermal features
are quite distinct of the usual thermal thermodynamics effects. For example, it is a highly correlated state with EPR-type correlations,\cite{pringle} and not a thermal
uncorrelated state as the equilibrium states in statistical thermodynamics. Some of the most feasible proposals are related to nonadiabatic conditions (i.e., capable of 
producing very rapid oscillations) for those cases in which
the nonadiabaticity parameter depends on the proper acceleration. It is only for this reason that an association with Unruh's effect is mentioned. A direct detection of
the scalar vacuum spectra has not been proposed so far. It requires noninertial propagation of a source such as a dislocation or a vortex through the 
corresponding phonon medium. 
For more details the interested reader can look at the extended version in electronic preprint form.\cite{extV}

%\subsection{Lists of items}
%\noindent
%Lists may be laid out with each item marked by a dot:
%\begin{itemlist}
% \item item one,
% \item item two.
%\end{itemlist}
%Items may also be numbered in lowercase roman numerals:
%\begin{romanlist}
%\item item one
%\item item two
%        \begin{alphlist}
%        \item Lists within lists can be numbered with lowercase
%              roman letters,
%        \item second item.
%        \end{alphlist}
%\end{romanlist}
%\newpage

\nonumsection{Acknowledgements}
\noindent
%\section*{Acknowledgment}

\noindent
The author very much appreciates the invitation of Dr. Pisin Chen to QABP3 and the kindness of the hosts at Higashi-Hiroshima.
He also is very grateful to Prof. Marlan O. Scully and people in his group for discussions on the `Unruh mazer' proposal. 

%This section should come before the References. Funding
%information may also be included here.

\nonumsection{References}
%\noindent
%References are to be listed in the order cited in the text. Use
%the style shown in the following examples. For journal names,
%use the standard abbreviations. Typeset references in 9 pt Times
%Roman.

%\appendix

%\noindent
%Appendices should be used only when absolutely necessary. They
%should come after the References. If there is more than one
%appendix, number them alphabetically. Number displayed equations
%occurring in the Appendix in this way, e.g.~(\ref{that}), (A.2),
%etc.
%\begin{equation}
%\mu(n, t) = {\sum^\infty_{i=1} 1(d_i < t, N(d_i) = n) \over
%\int^t_{\sigma=0} 1(N(\sigma) = n)d\sigma}\,. \label{that}
%\end{equation}
\end{document}